\documentclass{aa}  

\usepackage{amsmath,amssymb}
\usepackage[normalem]{ulem}
\usepackage{bm}
\usepackage{comment}
\usepackage{xcolor}
\usepackage{soul}
\usepackage[percent]{overpic}
\usepackage{overpic}
\usepackage{natbib}
\usepackage{amssymb,amsmath}

\usepackage{booktabs}

\newcommand{\unit}{1\!\!1}
\newcommand{\bma}{\begin{pmatrix}}
\newcommand{\ema}{\end{pmatrix}}

\newcommand{\bea}{\begin{eqnarray}}
\newcommand{\eea}{\end{eqnarray}}

\newcommand{\be}{\begin{equation}}
\newcommand{\ee}{\end{equation}}        
\newcommand{\ba}{\begin{eqnarray}}
\newcommand{\ea}{\end{eqnarray}}

\usepackage[T1]{fontenc}
\usepackage{ae,aecompl}

\begin{document}

\title{Pulsar current sheet \u{C}erenkov radiation}

   \author{Fan Zhang
          \inst{1}\inst{2}
          }
   \institute{Gravitational Wave and Cosmology Laboratory, Department of Astronomy, Beijing Normal University, Beijing 100875, China \\ \email{fnzhang@bnu.edu.cn}
         \and Department of Physics and Astronomy, West Virginia University, PO Box 6315, Morgantown, WV 26506, USA}
   \date{Received Oct 27, 2017; accepted }

  \abstract
{Plasma-filled pulsar magnetospheres contain thin current sheets wherein the charged particles are accelerated by magnetic reconnections to travel at ultra-relativistic speeds. On the other hand, the plasma frequency of the more regular force-free regions of the magnetosphere rests almost precisely on the upper limit of radio frequencies, with the cyclotron frequency being far higher due to the strong magnetic field. This combination produces a peculiar situation, whereby radio-frequency waves can travel at subluminal speeds without becoming evanescent. The conditions are thus conducive to \u{C}erenkov radiation originating from current sheets, which could plausibly serve as a coherent radio emission mechanism. In this paper we aim to provide a portrait of the relevant processes involved, and show that this mechanism can possibly account for some of the most salient features of the observed radio signals. }

\maketitle

\section{Introduction}
\begin{figure}[b!] 
  \centering
\begin{overpic}[width=0.95\columnwidth]{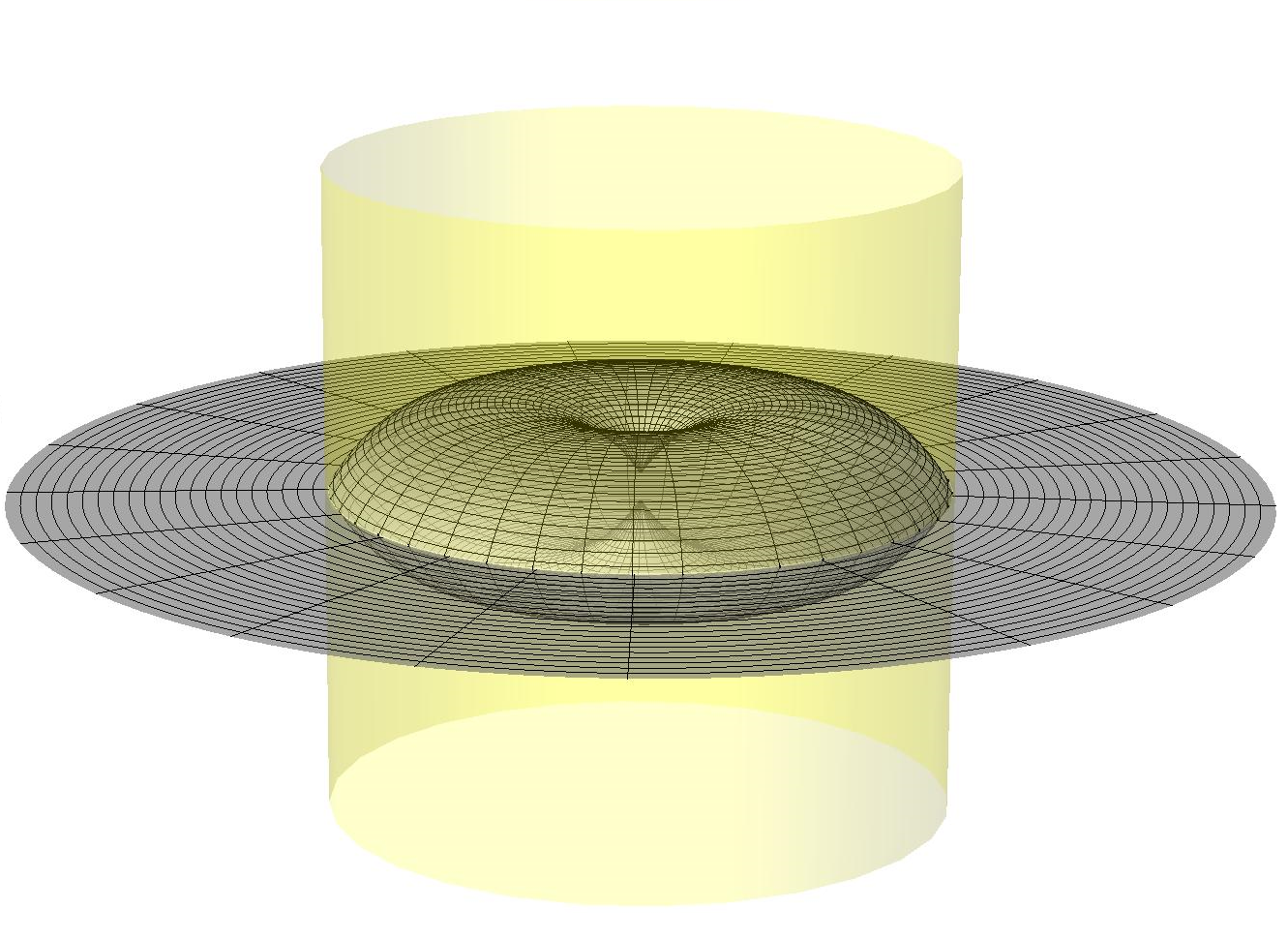}
\put(15,64){Light Cylinder}
\put(75,18){Equatorial CS}
\put(50,47){Separatrix CS}
\end{overpic}
  \caption{Current sheets (CSs) for an aligned rotator, whose cross-sectional shape matches that computed by \citet{2016ApJ...833..258G}. The yellow translucent surface represents the light cylinder, beyond which any co-rotating charged particles will have to travel superluminally and thus cannot exist. The CS outside of it resides on the equatorial plane, and those inside form a separatrix enclosure for the closed field line region. The concentrated currents flowing towards the neutron star at the centre of the figure along these sheets close the magnetosphere circuit (serve as the return current to those flowing out along open field lines), and provide the necessary boundary conditions that partition the open and closed field lines (allow the required discontinuities across the sheets, since open and closed lines have quite different toroidal components, see Eq.~\eqref{eq:Bphi} below).}
        \label{fig:CSPic}
\end{figure}

Pulsars have been observed to radiate into a wide range of electromagnetic wavebands, going from sub-megahertz up to as high as gamma-ray frequencies. Traditionally, much of the discussion of the plausible emission mechanisms have concentrated on the so-called gaps wherein the electric field along the magnetic field does not vanish. For example, higher frequency incoherent signals are proposed to originate in outer \citep{1986ApJ...300..500C}, slot \citep{1979ApJ...231..854A}, polar \citep{1978ApJ...225..226H} or inner annular \citep{2004ApJ...606L..49Q} gaps. 
Enabled by sophisticated magnetohydrodynamic or particle-in-cell simulations, some of the more recent studies accounting for a plasma-filled force-free magnetosphere have turned attentions to thin current sheets (CSs) (see Fig.~\ref{fig:CSPic} for a visual depiction) as alternative or additional sites of particle acceleration \citep{2014ApJ...795L..22C,2015ApJ...801L..19P,2016JPlPh..82b6302M}, and proposed for example, magnetic reconnections in them \citep{1996A&A...311..172L,2014ApJ...780....3U} as the source of high frequency emissions.
In this paper, we continue on along this path of exploring the roles played by the CSs, and we examine whether they could also be responsible for the less well understood (see e.g. \citet{2016JPlPh..82b6302M} for a critical review of the existing models) coherent radio emissions. 

\begin{table*}[tb] \label{tb:Params}
  \centering
  \label{tab:params}
  \begin{tabular}{cccc}
    \toprule
    Parameter & Symbol & Value & Comments\\
    \midrule    \midrule
    Magnetic field strength & $|{\bf B_e}|$ & $10^{8}\,$T & See \citet{2005AJ....129.1993M} for a detailed list \\
    &&& of individual pulsars. \\
    \midrule
        Rotation angular frequency & $\Omega$ & $2\pi \, \text{s}^{-1}$ & \\
    \midrule
        Neutron star mass & $M$ & $1.4\,\text{M}_{\odot}$ &  \\
    \midrule
        Neutron star radius & $R_*$ & $1.2\times 10^4\,\text{m}$ &  \\
    \midrule
        Goldreich Julian density & $n_{\rm GJ}$ & $10^{17}\,\text{m}^{-3}$ & \citet{Goldreich:1969sb}. This is the plasma density \\
    &&& if fully charge-separated. \\
    \midrule
        Plasma density in bulk & $n$ & $10^{19}\,\text{m}^{-3}$ & $n=\left(10^1-10^3\right) n_{\rm GJ}$ by \citet{2001ApJ...560..871H} \\
    &&& and \citet{2002ApJ...581..451A}.  \\
    \midrule
        Temperature & $T$ & $8\times 10^{6}\,\text{K}$ &  From persistent X-ray emissions \\ 
    &&& of polar cap regions \citep{2002ApJ...577..917K}. \\
    \midrule
        Average CS electron Lorentz factor & $\bar{\gamma}_{\rm CS}$ & $1000$ &  Numerical simulation by \citet{2015MNRAS.448..606C} \\
    \bottomrule
  \end{tabular}
  \caption{Typical values for normal pulsars (magnetic field strength $10^{7}-10^{9}$T) used as fiducial examples in this paper.}
\end{table*}

An essential characteristic of the CSs (distinguishing them from gaps for example) is that they tend to be very thin, because the zero resistivity force-free regions have a tendency to squeeze them, driving the volume current density and thus charged particle speeds to large values. Outside of the CSs, the plasma frequency $\omega_p$ for a neutron star magnetosphere is approximately in the tens to hundreds of gigahertz range (see Table~\ref{tb:Params} below for parameter values that go into this estimate), at the top end of radio frequencies. On the other hand, the cyclotron frequency $\omega_c$ is eight orders of magnitudes larger, instead of being much smaller as in the case with more familiar terrestrial plasma conditions, a fact that invalidates the intuition of the plasma frequency being a cutoff below which the waves become evanescent. Alternatively stated, the radio frequencies occupy a sweet spot of the pulsar magnetospheric environment, with the charged particles having sufficient time to respond to a passing radio wave (just like a simple pendulum will not move much under a driving force oscillating at frequencies higher than its natural frequency) and alter its dispersion relation, yet without being able to efficiently drain energy from it as their motions are severely constrained by the strong magnetic field. Consequently, at radio frequencies, the waves excited by those charged particles in a CS may well have phase speeds smaller than the speeds of the sourcing particles themselves, fulfilling the conditions for \u{C}erenkov radiation. Therefore, the detailed radio emission mechanism we propose is \u{C}erenkov\footnote{\u{C}erenkov processes (usually in conjunction with cyclotron radiations) have been evoked in maser type models \citep{1991MNRAS.253..377K,1999ApJ...512..804L,1999MNRAS.305..338L} (see also \citet{1994MNRAS.268..361A} for an exotic model of superluminal vacuum \u{C}erenkov emission from beyond the light cylinder). In this paper, we have applied the \u{C}erenkov radiation mechanism to a very different physical setup, where no growing instabilities are required and the source consists of an essentially steady flow of highly energetic particles within the CSs.}, 
which can robustly produce direct (able to escape the magnetosphere without requiring additional mode conversions) and broad-band (radio emissions expand over ten octaves \citep{2012hpa..book.....L}, without resonance peaks) non-thermal coherent radiation. Furthermore, the \u{C}erenkov radiation can be toggled between on and off states in a sharp binary fashion, which is convenient for explaining pulsar nulling. Specifically, if the plasma becomes depleted and the magnetosphere enters into a more strongly charge-separated quasi-stable state, the refractive index would drop towards unity, shutting down \u{C}erenkov radiation abruptly once the wave speeds exceed those of the sourcing particles. Such radio silence would occur concurrently to a starvation of the pulsar wind, explaining why pulsar nulling should be accompanied by a decline in the spindown rate, as is observed by for example, \citet{2006Sci...312..549K}, even though radio emission only constitutes a very small fraction of the total energy flux.

We present a more quantitative discussion of the CS \u{C}erenkov radiation (CSCR) model in Sect.~\ref{sec:Quant} and have examined its compatibility with salient features of the observed radio signals in Sect.~\ref{sec:Obs}, before concluding in Sect.~\ref{sec:Conc}. Concretely, we have substituted numerical values into derivations when assessing the relative importance of various terms, and also when comparing with observations. We collect in Table~\ref{tb:Params} the typical parameter values (in SI units, which are adopted by much of the plasma literature) applicable to normal pulsars. All the parameters and formulae below are presented in the pulsar frame. 

Because the misalignment between the spin and magnetic axes is not a prerequisite for the emission mechanism, we have assumed aligned rotator for tractability. { Nevertheless, we note that the CS structure does undergo noticeable quantitative changes when the rotator becomes oblique, as the displacement currents smooth out some of the sharper features of the CSs. This dependence on inclination angle is visible in, for example, the numerical simulations performed by \citet{Spitkovsky:2006np,2010ApJ...715.1282B} (see Figs.~1-3 in the latter reference). While the general morphology of two separatrix CSs connecting onto an equatorial one (see Fig.~\ref{fig:CSPic}) remains unchanged (unless the inclination angle reaches the extreme value of $\pi/2$), the so-called \verb!Y!-point where the three come together thickens and the separatrix CSs become weaker and wider. In principle, these should have quantitative impacts on the strength of the CSCR, as the width of the CSs is related to the charged particle speeds (see Sect.~\ref{sec:CS} below) and the vicinity of the \verb!Y! point is a prime location for particle acceleration \citep{2015MNRAS.448..606C}. However, it is currently difficult to gauge exactly how large their influences would be, because the CS details for the oblique rotator depend on realistic treatments of relevant factors such as the resistivity in the bulk of the magnetosphere. We anticipate interesting utilities if it turns out that the CSCR generated radio signals carry clean signatures of the inclination angle within its amplitude (e.g. a weakening due to the degradation of the separatrix CSs). }

\section{The emission mechanism} \label{sec:Quant}
In this section, we collect the various relevant ingredients and feed them into the conditions for \u{C}erenkov radiation, yielding emission parameters that will be used in Sect.~\ref{sec:Obs} to help interpret observations within the CSCR context. We have mostly followed the notation in \citet{1966RPPh...29..623M}, where boldface fonts denote vectors and tensors, but use $\otimes$ to denote tensor product. 

\subsection{The current sheets} \label{sec:CS}
The \u{C}erenkov radiation is predicated on the charged particles in the CSs moving at very high speeds, which is indeed demonstrated by sophisticated particle-in-cell simulations. For example, Fig.~9 in \citet{2015MNRAS.448..606C} shows the Lorentz factors achieved for electrons and positrons in the CSs (fed by energies released during magnetic reconnections). The electrons reach average Lorentz factors of $\bar{\gamma}_{\rm CS} \sim 1000$ (speed of $\bar{v}_{\rm CS} = 0.9999995c$) in the separatrix CSs (while the positrons reach similar speeds in the equatorial CS outside of the light cylinder), with their distribution (Fig.~10 of \citet{2015MNRAS.448..606C}) exhibiting a tail going up to as high as $3000$ or $\bar{v}_{\rm CS} = 0.99999995c$. 

{In fact, even these impressive $\bar{\gamma}_{\rm CS}$  values in the thousands may be underestimates (perhaps due to the choice of boundary conditions imposed on the inner boundary of the computational domain of \citet{2015MNRAS.448..606C}, which would not affect the conclusions of that work as it is mostly concerned with the conversion of Poynting flux into particle kinetic energy, via magnetic reconnection), as additional acceleration mechanisms might be active. Specifically, the energy available for accelerating electrons in the region close to the stellar surface can be estimated by computing the vacuum potential difference between the poles and the equator, which turns out to be able to boost the electrons to a Lorentz factor of around $\gamma_0=(\Omega R_*/c)(e |{\bf B}_e| R_*/mc)$ (see \citet{2015ApJ...801L..19P,2017arXiv170704323P}, but note we are in SI units thus have an extra factor of $c$). This $\gamma_0$ is of the order of $10^{11}$ for our fiducial parameters (using the magnetic field strength near the star), and indicates that there is much headroom for even greater $\bar{\gamma}_{\rm CS}$, which would further enhance the likelihood of \u{C}erenkov radiation occurring. In short, the rotation-induced electric field, if not perfectly shielded within the CSs, can provide an additional and possibly even greater accelerating force (more active closer to the star where the magnetic field is stronger) besides the magnetic reconnections (most active near the \verb!Y! point), and our fiducial value of $\bar{\gamma}_{\rm CS} \sim 1000$ should best be regarded as a conservative lower bound.}

Intuitively, such extreme values are not surprising, as ultra-relativistic speeds are consistent with, as well as contribute positively to the maintenance of quasi-steady (stable in the direction along the magnetic field) thin CSs. 
The most obvious fact regarding consistency emerges when we recall that the purpose for the existence of the CSs is to divide two very different magnetic regimes. Specifically, we have that in the geometrized units \citep{2017A&A...598A..88Z}
\begin{align} \label{eq:Bphi}
B^{\tilde{\phi}\, (c)}_e \sim 0\,, \quad B^{\tilde{\phi}\, (o)}_e \sim \frac{\sqrt{2}\mu \Omega^2\csc^2\tilde{\theta}}{\tilde{r}^{3/2}\sqrt{2\tilde{r}-1}}\,,
\end{align}
where $(o)$ and $(c)$ stand for the open and closed field line regions respectively, and $\mu$ is the magnetic dipole moment. The quantity $\tilde{r}$ is the normalized radius so $\tilde{r}=1$ on the stellar surface, and $\tilde{\theta}$ is the location angle against the magnetic axis (coinciding with the spin axis for our aligned rotator). Such an abrupt disparity means the derivative of the toroidal magnetic field across a very thin CS would be extremely large, requiring a large poloidal current density (thus high charged particle speeds, {but also a high charge density when the speeds saturate at close to the speed of light}) to sustain. 
This intuition can be (crudely) quantified by evoking the equilibrium Harris CS \citep{1962NCim...23..115H}, extended to the relativistic case by \citet{1966PhFl....9..277H} and extensively utilized in later pulsar CS studies such as \citet{2013MNRAS.434.2636P}. The central prediction is that the magnetic field variation across a CS satisfies
\bea
{\bf B}_e=\bar{\bf B}_e + \frac{\Delta {\bf B}_e}{2} \tanh(d/\lambda)\,, 
\eea
where $d$ is the distance to the mid-plane of the CS, $\bar{\bf B}_e$ is the average magnetic field near the sheet and $\Delta {\bf B}_e$ the jump in magnetic field across it. The thickness of the sheet is given by {(see Appendix A of \citet{1966PhFl....9..277H})}
\bea \label{eq:Ampere}
\lambda = \frac{|\Delta {\bf B}_e|}{\mu_0}\frac{1}{4 e n_0 U}\,, 
\eea
{where $U$ is the magnitude of the four velocity $U^a =\bar{\gamma}_{\rm CS}(1,\bar{\bf v}_{\rm CS})$, $e$ denotes charge and $n_0$ is the number density (overhead bar signifies averaging).} The sheet is thus more compressed when the charged particles move quickly. Equation~\eqref{eq:Ampere} is essentially just the Amp\`{e}re's law, so although the actual CS may have a different detailed internal structure (e.g. not necessarily of a $\tanh$ functional form), Eq.~\eqref{eq:Ampere} will remain a valid quantitative illustration that thin current sheets demand high particle speeds. 

Regarding the maintenance of a quasi-steady-state, we note that the ideal infinite conductivity condition in the force-free regions of the magnetosphere would try to squeeze the CSs to become infinitesimally thin, forcing $\bar{\gamma}_{\rm CS}$ to diverge, leading to a large Reynolds number and instabilities, unless a counteracting dissipative process exists to introduce an effective resistivity that keeps $v_{\rm CS}$ close to some finite equilibrium value. We posit that the \u{C}erenkov radiation reaction constitute an ideal candidate mechanism to accomplish this task, which is then somewhat similar in intended functionality to dissipation through unbridled particle acceleration often assumed for the gaps \citep{2016JPlPh..82b6302M}, but which turns out to be inefficient because violent instabilities still develop that force the gaps to become temporally highly variable \citep{2005ApJ...631..456L}. In contrast, the \u{C}erenkov radiation reaction is more genuinely dissipative (excess energy propagates away as waves; alternative kinetic mechanisms such as traditional dissipation by collisions tends to randomize but not remove the energy, thus instabilities may take on a different form but persist), and should result in the CSs being more stable. 

\subsection{The magnetospheric environment} \label{sec:Params}
In this section we examine wave propagation in the more regular parts of the magnetosphere, through which the waves sourced by the CSs must traverse. Many extensive studies on such waves exist in literature, assuming different magnetospheric conditions. For convenience though, we carried out an ab initio computation in the next subsection so we can extract intermediate results that are useful when examining whether the \u{C}erenkov radiation conditions are satisfied. We note however, the fact that subluminal waves exist is not a new result, nor is it sensitive to the assumptions we adopt in this work. 

For tractability and brevity, it is necessary to make simplifying assumptions that are reasonable for the pulsar magnetospheres. We enumerate them in this section, and note that they inevitably restrict the variety of the \u{C}erenkov waves that can be generated. In particular, our discussion applies to the normal pulsar population with magnetic field strength around $10^{8}$T, and not the particularly young or old ones for which different or more diverse behaviours are possible. Our assumptions are:
\begin{enumerate}
\item Because the radio frequencies can be of the same order as the plasma frequency, we are not in the magnetohydrodynamic regime \citep{2015bps..book.....C}. In particular, force-free electrodynamics, although valid for describing the stationary background configurations of the magnetospheres, are not applicable when studying the radio frequency waves travelling within. We will therefore rely on the more fundamental kinetic equations instead. 

\item As the radio emission region is not confined to be immediately abutting the stellar surface or the light cylinder, we assume that the emission mechanism is not sensitive to the boundary conditions they might impose. 

\item The collisions between particles are not included, as per usual when studying the force-free (rarefied plasma) pulsar magnetospheres (see e.g. \citet{1986ApJ...302..120A}). This means no consideration is given to the annihilations between electrons and positrons. We also have not considered pair production, which is assumed to be mostly limited to confined regions (we are presently considering regular bulk regions of the magnetosphere). Therefore, the Vlasov equation derived from conservation of particle numbers is assumed valid. 

\item The plasma is assumed `cold', with $k_{\rm B} T\ll mc^2$ ($T$ stands for temperature, and $k_{\rm B}$ the Boltzman constant), since observationally, the plasma temperature is estimated to be $k_{\rm B}T/(mc^2) \sim 10^{-3}$ (see Table~\ref{tb:Params}). This is also consistent with the plasma being ideal or force-free (see e.g. Ch.~4 of \citet{2015bps..book.....C}), and saves us the uncertainty of a velocity distribution function.  

\item Low temperature in a plasma sometimes leads to the necessity for quantum corrections, so we checked that this is not needed in our case. Using the data in Table~\ref{tb:Params}, we can estimate the de Broglie wavelength at $\hbar/(3mk_BT) \sim 6\times 10^{-12}$m, which is indeed much smaller than $n^{-1/3}\sim 5\times 10^{-7}$m. In addition, the magnetic field strength is not in excess of the quantum electrodynamic threshold of $m^2 c^3/(he)\sim 4.4 \times 10^{9}$T, so the vacuum does not become birefringent. 

\item Numerically, the bulk regions (outside of the CSs) of Fig.~9 in \citet{2015MNRAS.448..606C} show that the collective streaming speeds of the charged particles are small to moderate. Analytically, one may apply the rotating dipole simplification and compute the E-cross-B or electric drift velocities of the charged particles, whose magnitudes turn out to be on par with the co-rotation speed \citep{2016JPlPh..82b6302M}, and are thus non-relativistic for our proposed emission region at much lower altitudes than the light cylinder. We therefore set the streaming speeds to zero below (also note that their being non-relativistic is implicitly assumed in the usual magnetohydrodynamic -- not utilized here -- studies of the magnetospheres). This assumption is in fact different from some earlier kinetic studies on pulsar magnetospheres assuming ultra-relativistic streaming speeds in the bulk region (see e.g. \citet{1986ApJ...302..120A}), but the differences are mainly attributable to aberration or beaming and a Doppler shift in frequency \citep{2016JPlPh..82b6302M} (assuming that the thermal motion remains non-relativistic in the plasma rest frame). 

\item As the gravitational force is much smaller than electromagnetic forces in the pulsar magnetosphere \citep{Goldreich:1969sb} (using parameters in Table~\ref{tb:Params}, the charged particles need to move slower than $10^{-7}\text{m}\,{\text s}^{-1}$ for the gravitational force on the surface of the star to be comparable to the Lorentz force), we ignore it and thus only electromagnetic forces appear in the Vlasov equation. As an aside, we also mention that the Coriolis force in the co-rotating frame (we do not use this frame in our computations) would not become dominant unless the speed of the particles differs from $c$ by less than $10^{-37}$ in relative terms. It may however generate more subtle secondary effects when applied to the much more fast-moving particles inside of the CSs, see Sect.~\ref{sec:TempProf} below. 

\item We have assumed similar densities for the electrons and positions, as plasma creation is dominated by pair production as opposed to extraction from the star, and charge separation is low (see Table~\ref{tb:Params}).

\end{enumerate}
We note that although this long list of simplifying assumptions were adopted to allow for a more quantitative and less cluttered illustration of the essential properties of the emitted waves, the underlying condition for \u{C}erenkov radiation to occur is satisfied more generically. For example, allowing for high temperatures and/or relativistic relative streaming, \citet{1998MNRAS.293..447L} computed the refractive index for a pair plasma, which also contains parameter ranges allowing for subluminal wave propagation. 

A word of caution is also in order. The discussions in the following sections aim to establish that the basic conditions for \u{C}erenkov emission are met, and to compute the rudimentary geometric features of the radiation. However, the real waves may be more complicated than the linearized formalism could properly account for, as they crowd onto a shock front and become concentrated. 
Even the fully non-linear equations may only be satisfied in a distributional sense (i.e. true discontinuities may exist to necessitate a generalization of the definition of derivatives, the waves are then only weak solutions to the propagation equations). Therefore, we expect that an accurate account of the finer details, such as complications seen in the polarization states (which are also sensitive to a myriad of other propagation effects \citep{2008AIPC..983...29L}), would emerge only with more sophisticated treatments in the future.

\subsection{The phase velocities}
The starting point towards the dispersion relations is the Vlasov equation (see e.g. \citet{1966RPPh...29..623M})
\bea \label{eq:Vlasov}
\frac{\partial f^{\pm}}{\partial t}+ {\bf w}\cdot \frac{\partial f^{\pm}}{\partial {\bf x}}+\frac{q^{\pm}}{m}\Big[{\bf E}_{e}+{\bf E}+{\bf w} \times ({\bf B}_{e}+{\bf B})\Big]\frac{\partial f^{\pm}}{\partial {\bf w}} = 0\,,
\eea
where ${\bf x}$ is the location, ${\bf w}$ the velocity of a particle, and $f^{\pm}({\bf x},{\bf w})$ describes the distribution in phase space for electrons ($-$ sign) and positrons ($+$ sign). The quantity ${\bf B}_{e}$ (${\bf E}_{e}$) denotes the background magnetic (electric) field, while ${\bf B}$ (${\bf E}$) is the magnetic (electric) field associated with the waves. We can further define the mean value (over velocity) of any arbitrary quantity $A$ as 
\bea
\langle A \rangle^{\pm} \equiv \frac{1}{n^{\pm}} \int A f^{\pm} d{\bf w}\,, \quad \text{where} \quad
n^{\pm} \equiv \int f^{\pm} d{\bf w}\,. 
\eea
A tower of equations can be constructed out of Eq.~\eqref{eq:Vlasov} by setting $A$ to higher and higher tensor products of ${\bf w}$ ($1$, ${\bf w}$, ${\bf w}\otimes {\bf w} \cdots$), essentially turning that equation for the distribution function into those for its moments. This tower of equations is chained, with quantities appearing in a lower level one being determined by the next level up. The chain can be terminated, for example, at the second level by adopting the fully adiabatic hypothesis, giving an error at the order $(V^{\pm}_T/v_{\phi})^4$, where $v_{\phi}$ is the phase speed of the wave under consideration, and $V^{\pm}_T$ is the thermal velocity, which is assumed negligible in our present consideration (see Item 4 of Sect.~\ref{sec:Params}). 
One then works down from the top of the tower towards the collective flow velocity ${\bf v}^{\pm}\equiv \langle {\bf w} \rangle^{\pm}$ and thus the current, which can be substituted into the Maxwell's equations to yield a coupled set of linearized equations (the wave equation)
\bea \label{eq:EqE}
{\bf W}_E {\bf E} = 0\,,
\eea
where
\bea \label{eq:Det}
{\bf W}_E = {\bf k}\otimes {\bf k}-k^2 \unit+\frac{\omega^2}{c^2}\left(\unit+\frac{{\bf T}}{i\omega \epsilon_0}\right)\,,
\eea
and which must satisfy 
\bea \label{eq:Cond}
\det {\bf W}_E =0
\eea
for Eq.~\eqref{eq:EqE} to admit non-trivial solutions. The quantity ${\bf T}$ is the conductivity tensor, and the dielectric tensor is further given by $\bm{\epsilon}=(\unit+{\bf T}/(i\omega \epsilon_0))$. Condition \eqref{eq:Cond} then determines the dispersion relations for all the possible modes of propagation. Aside from determining the phase and group velocities, the dispersion relations also reveal instabilities and damping in modes through the imaginary parts of the frequencies. Further solving for ${\bf E}$ from Eq.~\eqref{eq:EqE} will reveal whether the wave is transverse (${\bf E}\cdot {\bf k}=0$), and what is its polarization. For example, for a transverse wave, let ${\bf e_1}$ and ${\bf e_2}$ be two orthogonal basis vectors transverse to ${\bf k}$, then the wave is linearly polarized if $\xi\equiv ({\bf E}\cdot {\bf e_1})/({\bf E}\cdot {\bf e_2}) \in \mathbb{R}$, circularly polarized if $\xi =\pm i$, elliptically polarized if $\xi$ is some other arbitrary imaginary number, and mixed if both $\Re \xi \neq 0$ and $\Im \xi \neq 0$.  

The expression for ${\bf W}_E$ is given by Eq.~40 of \citet{1966RPPh...29..623M} (note there is a typo in the bottom right corner in that reference)
\begin{align} \label{eq:WE}
&{\bf W}_E = \notag \\
&\bma 
-k^2 \cos^2\theta + \frac{\omega^2}{c^2} S & i \frac{\omega^2}{c^2} D & k^2 \sin\theta \cos\theta \\
- i \frac{\omega^2}{c^2} D & -k^2 + \frac{\omega^2}{c^2} S & 0 \\
k^2 \sin\theta \cos\theta  & 0 & -k^2 \sin^2 \theta + \frac{\omega^2}{c^2} P
\ema\,,
\end{align}
in an adapted co-ordinate system where the magnetic field is along the third axis. The angle $\theta$ is that between ${\bf B}_e$ and ${\bf k}$, with the first axis chosen so ${\bf k}$ lies on the plane spanned by the first and third axes. The auxiliary quantities appearing in Eq.~\eqref{eq:WE} are
\bea
S\equiv 1-\frac{2\omega^2_p}{\omega^2 -\omega_c^2}\,,\quad 
P\equiv 1-\frac{2\omega^2_p}{\omega^2}\,, \quad D\equiv 0
\eea 
for the pair plasma with equal number densities (note $D\neq 0$ when $n^+\neq n^-$, which must be true to some extent for the plasma to be able to enforce force-freeness -- the two densities differ by around $1\%$ given the parameters shown in Table~\ref{tb:Params}, but this small difference is neglected in the present discussion for simplicity, as per Item 8 in Sect.~\ref{sec:Params}; resurrecting $D$ by either admitting this or other differences between the electrons and positrons will produce elliptical or circular polarizations), while the plasma and cyclotron frequencies are 
\bea
\omega_p \equiv \sqrt{\frac{n (q^{\pm})^2 }{m\epsilon_0}}\,, \quad \omega_c \equiv \frac{|q^{\pm} {\bf B}_e|}{m}\,.
\eea
Equation~\eqref{eq:Cond} then dictates that the refractive index $\eta \equiv kc/\omega$ must satisfy the equation
\bea \label{eq:refEq}
A \eta^4 - B \eta^2 + C = 0 \,,
\eea
where
\begin{align}
A &\equiv S \sin^2 \theta + P \cos^2\theta\,, \quad 
C \equiv P S^2 \,,
\notag \\
B &\equiv S^2 \sin^2 \theta + PS (1+\cos^2\theta)\,.
\end{align}

\begin{figure}[tb] 
  \centering
\begin{overpic}[width=0.43\columnwidth]{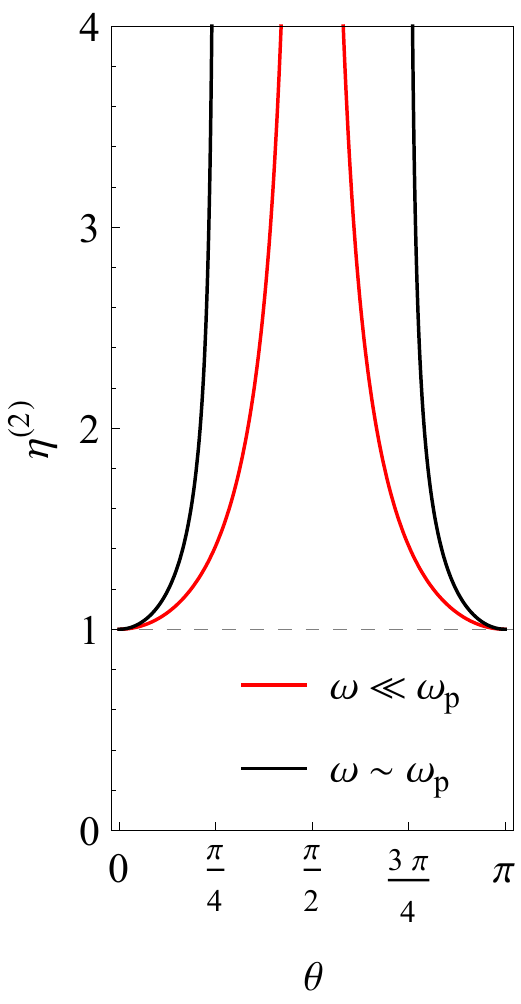}
\put(45,2){(a)}
\end{overpic}
\begin{overpic}[width=0.47\columnwidth]{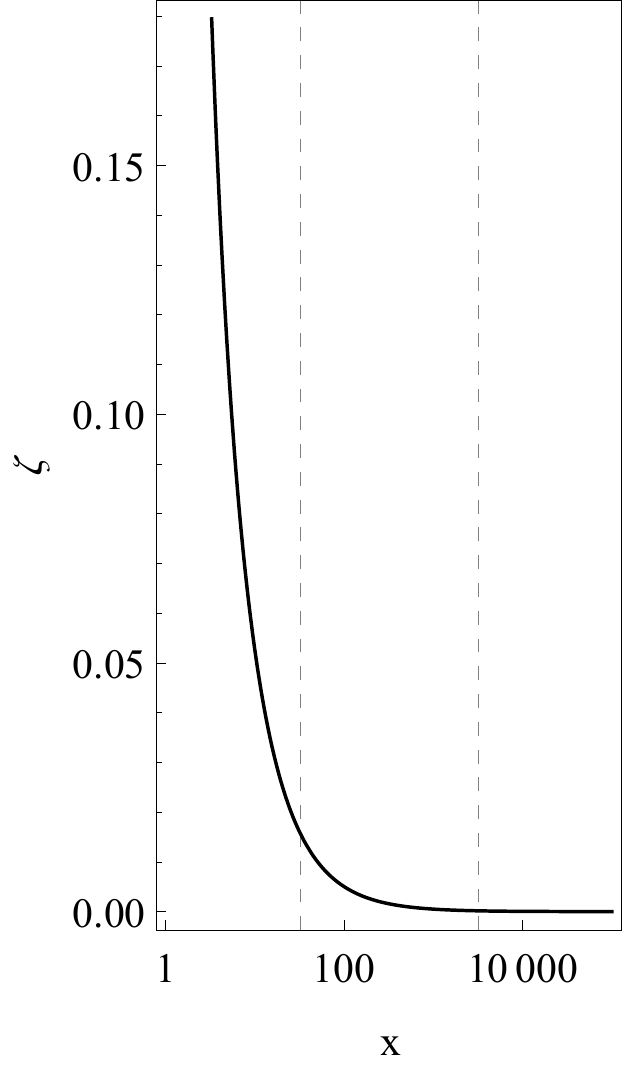}
\put(50,2){(b)}
\end{overpic}
  \caption{(a): Dependence of $\eta^{(2)}$ on $\theta$ for the two regimes of low and high radio frequencies. The divergences will be smoothed by collisions and finite temperatures (our cold plasma treatment becomes invalid when the thermal speed of the particles exceed the phase speed $c/\eta^{(2)}$ of the waves, so corrections will always set in at some stage for a diverging $\eta^{(2)}$) \citep{1966RPPh...29..623M}. (b): The dependence of $\zeta$ in the polarization vector \eqref{eq:Polar} on $x$, plotted with $y/x=\omega_p^2/\omega_c^2$ as given by Table~\ref{tb:Params}. The locations of $\omega = 2\pi \times 5$GHz and $2\pi\times 0.5$GHz are marked out with vertical dashed lines. }
        \label{fig:zeta}
\end{figure}

Defining $x=\omega_p^2/\omega^2$ and $y=\omega_c^2/\omega^2$, the solutions for Eq.~\eqref{eq:refEq} can be written as 
\begin{align} \label{eq:Eta}
\eta^{(1)} =& \sqrt{1+\frac{2x}{y-1}} \,, \notag \\
\eta^{(2)} =& \sqrt{\frac{(2 x-1) (2 x+y-1)}{x y \cos (2 \theta )+x (y-2)-(y-1)}} \,.
\end{align}
We note that $\eta^{(2)}$ becomes identical to $\eta^{(1)}$ when $\theta=0$. For oblique angles, while $\eta^{(1)}$ is always very close to unity (differs from it by around one in $10^{16}$ using parameter values in Table~\ref{tb:Params}, so our fiducial $v_{\rm CS}$ is insufficient for \u{C}erenkov radiation into this branch), $\eta^{(2)}$ can be substantially larger, whose angular dependence becoming $\sec\theta$ when $\omega \ll \omega_p$ and $\sqrt{1/\cos2\theta}$ when $\omega\sim \omega_p$ (see Fig.~\ref{fig:zeta} (a)). It is thus branch $(2)$ that we will mostly concentrate on.  

The solution to Eq.~\eqref{eq:EqE} corresponding to $\eta^{(2)}$ is (while for $\eta^{(1)}$ is ${\bf E}^{(1)} \propto (0,1,0)^T$)
\bea \label{eq:Polar}
{\bf E}^{(2)} \propto \left(1,0,\zeta \tan\theta \right)^T \equiv \left(1,0,\frac{2x-1+y}{(2x-1)(y-1)}\tan\theta\right)^T\,.
\eea
One can also compute the conductivity tensor ${\bf T}$ with Eq.~\eqref{eq:Det}, from which one can infer that the alternating current density is given by 
\bea \label{eq:PolarCurrent}
{\bf j}^{(2)} = {\bf T}\,{\bf E}^{(2)}=\left(-\frac{2 i \omega  \omega_p^2 \epsilon_0}{\omega ^2-\omega_c^2},\,0,\,-\frac{2 i \zeta  \omega_p^2 \epsilon_0 \tan \theta}{\omega }\right)\,.
\eea
The quantity $\zeta$ quickly drops towards zero with increasing $x$ (see Fig.~\ref{fig:zeta} (b), and note $\zeta = 1/(2x-1)$ in the $y\rightarrow \infty$ limit), so both ${\bf E}^{(2)}$ and ${\bf j}^{(2)}$ are essentially along the $(1,0,0)^T$ direction when $\omega \ll \omega_p$, orthogonal to ${\bf B}_e$. 

\subsection{The \u{C}erenkov radiation}
The conditions for \u{C}erenkov radiation to occur is relatively simple, and laid out in for example, \citet{1967RaSc....2..703S}, they are (note there is a misnomer in that reference, $\omega/k_{\parallel}$ is not a component of the phase velocity, but is larger than it in magnitude)
\bea
\frac{\omega}{v_{\rm CS}} = k_{\parallel}\,, \quad \text{and} \quad \Im k_{\bot} = 0\,, 
\eea
where $k_{\parallel}$ is the component of the wavenumber along the direction of source particle motion, and $k_{\bot}$ the one transverse to it. These expressions are the same for either electrons or positrons, and the second equation is satisfied as long as the phase speed of the wave is smaller than the speed of the source particle. For the first equation (which is a selection rule arising from the need to balance the exponential factors in the Fourier-transformed Maxwell's equations), we note that because the charged particles flow along the magnetic field lines, we have $\tan\theta=k_{\bot}/k_{\parallel}$, 
and thus 
\bea \label{eq:Direction}
\cos\theta\,\eta(\theta,\omega)=\frac{c}{v_{\rm CS}}\,,
\eea
that must be satisfied at the emission site (not necessarily during subsequent propagation). 
When the medium was isotropic, we recovered the familiar propagation direction $\cos\theta=c/(\eta v_{\rm CS})$ for a \u{C}erenkov wave. In our present context however, $\eta$ is in general a function of $\theta$, and so the propagation direction is slightly more complicated (when invoking the Huygens-Fresnel principal to intuit the \u{C}erenkov wave as being a shock front, the secondary wavefronts are no longer circular for us). For any given radio frequency $\omega$ under examination, Eq.~\eqref{eq:Direction} becomes an equation for $\theta$ that we can solve to ascertain the dominant \u{C}erenkov phase (not group) velocity direction. We note $v_{\rm CS}$ is not given by a delta function, so $\theta$ has a spread around the dominant value, which gives the emission cone some thickness (see below and Sect.~\ref{sec:Obs}). 

\begin{figure}[tb] 
  \centering
\begin{overpic}[width=0.95\columnwidth]{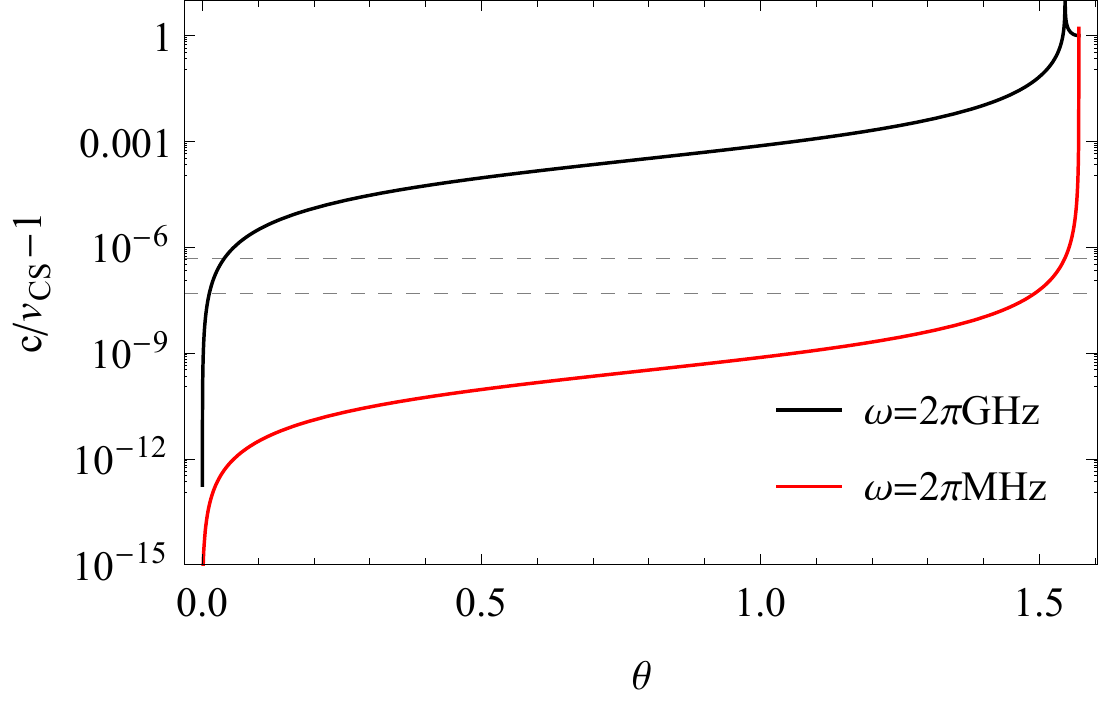}
\end{overpic}
  \caption{Left hand side of Eq.~\eqref{eq:Direction} (minus $1$) plotted as solid curves for the observational (non-angular) frequencies of $1$MHz and $1$GHz. The right hand side of the equation (minus $1$) for $v_{\rm cs} =0.9999995$c (upper) and $v_{\rm cs} =0.99999995$c (lower) are plotted as the dashed horizontal lines, so the solutions to Eq.~\eqref{eq:Direction} are the intersection points between solid and dashed lines. }
        \label{fig:Direction}
\end{figure}

\begin{figure}[tb] 
  \centering
\begin{overpic}[width=0.95\columnwidth]{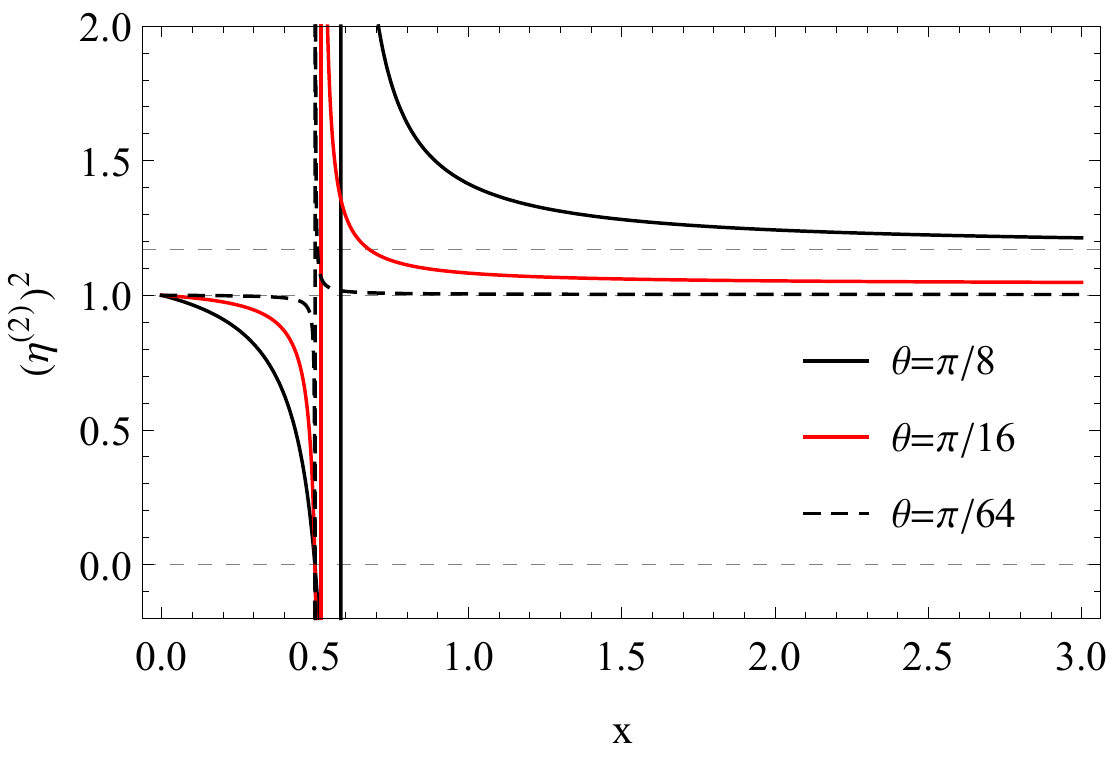}
\end{overpic}
  \caption{Expression \eqref{eq:Eta} (the plot for \eqref{eq:Propagation} is indistinguishable) as a function of $x$, plotted for fixed $\theta = \pi/8$, $\pi/16$ and $\pi/64$. The horizontal dashed lines are at $(\eta^{(2)})^2=0$, $1$, and $\sec^2\pi/8$. When $\omega \ll \omega_p$ or $x\gg 1$, $(\eta^{(2)})^2$ asymptotes to $\sec^2\theta$, while when $\omega \gg \omega_p$ or $x\ll 1$, $(\eta^{(2)})^2$ becomes unity regardless of the angle $\theta$. Note Eq.~\eqref{eq:Direction} only needs to be satisfied at the emission site, and determines the central phase velocity direction, the same $\theta$ is then kept by the wave as it propagates. So as the wave rises in altitude, the refractive index that the wave experiences at new sites along its trajectory can be estimated by simply tracing along one of the curves in this figure to the left ($\omega_p$ thus $x$ drops with increasing altitude, as the plasma becomes more rarefied). }
        \label{fig:Eta2Sqr}
\end{figure}

The solutions to Eq.~\eqref{eq:Direction} inherit a strong $\omega$ dependence (see Fig.~\ref{fig:Direction}) from $\eta^{(2)}$, and the group velocities of the emitted waves follow the magnetic field lines. To see this, we begin by noting that the expression \eqref{eq:Eta} for $\eta^{(2)}$ simplifies into (cf. \citet{2016JPlPh..82b6302M} Eq.~5.2, but note that we have extra factors of two arising from explicitly summing over particle species in a pair plasma, while some literature prefers an effective-one-specie approach and have it absorbed by $\omega_p$)
\bea \label{eq:Propagation}
\left(\eta^{(2)}\right)^2=\frac{k^2 c^2}{\omega^2} \approx \frac{2\omega_p^2-\omega^2}{2\cos^2 \theta\,\omega_p^2-\omega^2}\,,
\eea
in the infinite magnetic field ($y\rightarrow \infty$) limit. When $\omega_p$ is appreciably larger than $\omega$ (i.e. $x\gg 1$), the right hand side of Eq.~\eqref{eq:Propagation} becomes approximately $\sec^2\theta$ (see also discussion below Eq.~\ref{eq:Eta}), and the dispersion relation reduces to simply $\omega=\pm ck_{\parallel}$, thus $\nabla_{\bf k}\omega$ is along the magnetic field. In other words, we are essentially looking at the Alfv\'en wave (with a small amount of Langmuir wave blended in; while $\eta^{(1)}$ describes the electromagnetic mode) generated mostly in the dense plasma region at low altitudes (as $n$ drops towards zero at higher altitudes, $\eta^{(2)}$ -- as well as $\eta^{(1)}$ -- would approach unity, quenching \u{C}erenkov radiation) above the polar caps. As the Alfv\'en waves propagate to higher altitudes, they tunnel through a thin evanescent layer ($\eta^{(2)}$ briefly turn imaginary, see Fig.~\ref{fig:Eta2Sqr} and note that only the segment below the lowest horizontal dashed line is evanescent), and emerge as a superluminal (in phase velocity only) ordinary wave, which is no longer bound to the magnetic field lines (the term containing $\theta$ in the denominator of Eq.~\eqref{eq:Propagation} becomes sub-dominant, so the dispersion relation and thus the propagation properties of the wave become isotropic) and can escape the magnetosphere as simple vacuum waves when $n$ thus $\omega_p$ drops towards zero. A similarly direct (no need for scattering or mode coupling to hand over the energy to a different branch of the dispersion relations, which tends to be inefficient and thus pose a difficulty for many pulsar radio emission mechanisms \citep{2000ASPC..202..417U}) escape had been noted by \citet{1999JPlPh..62..233M,1999ApJ...521..351M,2016JPlPh..82b6302M} for a different type of wave (Langmuir) existing in a different environment (relativistic thermal plasma), produced by a different mechanism (beam instability).

Such a scenario leads to the radio signal being concentrated on a narrow hollow cone. Hollow because the radiations are generated only near the CSs, thus initially track the last open field lines (LOFL), and not spread out across the entire open-field-line region; narrow because the opening angle is set by the nearly vertical directions of the LOFL above the polar caps. For a more quantitative illustration then, it is beneficial to study the geometry of these lines more closely. Starting from the results of \citet{2016ApJ...833..258G} for the general relativistic force-free magnetosphere, which gives the non-vanishing components of the covariant Faraday tensor as 
\begin{align}
&F_{t\tilde{r}} = \Omega \frac{\partial \psi}{\partial \tilde{r}}\,,\quad  
F_{t\tilde{\theta}} = \Omega \frac{\partial \psi}{\partial \tilde{\theta}} \,,\quad 
F_{\tilde{\phi} \tilde{r}} = -\frac{\partial \psi}{\partial \tilde{r}} \,,\notag \\
&F_{\tilde{\phi} \tilde{\theta}} = -\frac{\partial \psi}{\partial \tilde{\theta}}\,,\quad 
F_{\tilde{r}\tilde{\theta}} = \frac{\tilde{r}}{\pi\sin\tilde{\theta}}\frac{\mathcal{I}}{2\tilde{r}-1} \,,
\end{align}
where $2\pi\psi(\tilde{r},\tilde{\theta})$ is the total magnetic flux across a surface bound by the toroidal curve of constant $\tilde{\theta}$ and $\tilde{r}$, and 
\bea
\mathcal{I}^{(o)} = \pm 2\pi \Omega \psi\left( 2-\frac{\psi}{\psi_0}-\frac{1}{5}\left(\frac{\psi}{\psi_0}\right)^3\right)\,, \quad \mathcal{I}^{(c)}=0
\eea
are the total current through the same surface. For the CSs that are tangential to the field lines, the $\psi$ value on them is a constant $\psi_0\approx 1.23\mu \Omega$ (recall $\mu$ is the dipole moment). These information are sufficient to yield the toroidal component of the magnetic field as given by Eq.~\eqref{eq:Bphi}. The other components $B^{\tilde{r}}$ and $B^{\tilde{\theta}}$ are more complicated as they involve derivatives of $\psi$, so we need to know the situation away from the CSs. The ratio between them is constrained by the magnetic field being tangential to the cross section of the CS (vertical cut of Fig.~\ref{fig:CSPic}), given by
\begin{align} \label{eq:CSProfile}
\frac{\tilde{z}}{R_{\rm LC}} \equiv \frac{\tilde{r}\cos\tilde{\theta}}{R_{\rm LC}} \approx &-2.2 \left(\frac{\tilde{\rho}}{R_{\rm LC}}\right)^4+4.2 \left(\frac{\tilde{\rho}}{R_{\rm LC}}\right)^3
\notag \\ &
-3.6 \left(\frac{\tilde{\rho}}{R_{\rm LC}}\right)^2+1.5 \left(\frac{\tilde{\rho}}{R_{\rm LC}}\right)
+0.085\,,
\end{align}
(where $\tilde{\rho}\equiv \tilde{r}\sin\tilde{\theta}$, $\tilde{z}\equiv \tilde{r}\cos\tilde{\theta}$, and $R_{\rm LC} \sim 5\times 10^{4}$km is the distance from the rotation axis to the light cylinder) as obtained by fitting to data in \citet{2016ApJ...833..258G}, and so we have
\bea
\frac{B^{\tilde{z}\,(c/o)}}{B^{\tilde{\rho}\,(c/o)}}=\frac{\cos\tilde{\theta}(\partial\psi/\partial \tilde{\theta})+\tilde{r}\sin\tilde{\theta} (\partial\psi/\partial \tilde{r})}{\sin\tilde{\theta}(\partial\psi/\partial \tilde{\theta})-\tilde{r}\cos\tilde{\theta} (\partial\psi/\partial \tilde{r})}=\frac{d\tilde{z}}{d\tilde{\rho}}\,,
\eea
near the CS (the derivative on the right is to be evaluated using Eq.~\ref{eq:CSProfile}). This condition is however not enough to fix both $\partial\psi/\partial \tilde{\theta}$ and $\partial\psi/\partial \tilde{r}$, and is cumbersome to use. Instead, for illustration, we adopt an adjustable dipole\footnote{\label{ft:Dipole}The dipole is used as the overlap regime (intermediate between but away from the star and the light cylinder) solution in the matched asymptotic expansion procedure of \citet{2016ApJ...833..258G}. } approximation $\psi \approx (\mu/\tilde{r})\mathfrak{g}(\tilde{\theta})$, where $\mathfrak{g}(\tilde{\theta})$ is a function of the poloidal angle, with value $\sim \mathcal{O}(1)$, giving (prime indicates derivative)
\bea
B^{\tilde{\rho}\, (o)/(c)}_e \sim \frac{ \mu \left(\mathfrak{g}'+\mathfrak{g}\cot\tilde{\theta} \right)}{\sqrt{2}}\frac{\sqrt{2 \tilde{r}-1}}{\tilde{r}^{7/2}}\,.
\eea
Noting that $B^{\tilde{\phi}\, (c)}_e$ (Eq.~\ref{eq:Bphi}) picks up another factor of $\tilde{r}$ when converted to Cartesian co-ordinates and that $\tilde{r} \Omega \approx 1$ near the light cylinder, we see that when viewed from above down the magnetic axis, the LOFLs would look largely radial when close to the axis but sweep back strongly when further out towards the light cylinder (on the other hand, the closed field lines remain exactly radial throughout). 
%In particular, with exact dipole where $\mathfrak{g}(\tilde{\theta})=\sin^2\tilde{\theta}$, the field lines thread through the equatorial plane orthogonally, so $\mathfrak{g}'(\pi/2)=0$ (also note $\cot(\pi/2)=0$) and thus $B^{\tilde{\rho}\, (o)/(c)}_e =0$, and the LOFLs would look completely toroidal at the light cylinder (the CSs intersect the equatorial plane there). In reality however, the separatrix CSs do not follow dipole field lines exactly and do not intersect the equatorial plane orthogonally (see Fig.~\ref{fig:CSPic}), so $\mathfrak{g}'(\pi/2) \neq 0$ (the $\psi\propto 1/\tilde{r}$ assumption is also no longer strictly valid near the light cylinder) and open field lines will be able to penetrate the light cylinder. 
We do not know the precise form of $\mathfrak{g}(\theta)$, but for our interested region deep inside the light cylinder, it suffices to use the dipole expressions $\partial\psi/\partial \theta = 2\psi\cot\theta$ and $\partial\psi/\partial r = -\psi/r$, but replace $\psi$ with $\psi_0$ that's more accurately reflective of the actual $\psi$ value on the CS than the dipole expression. We have thus 
\begin{align} \label{eq:BPol}
B^{\tilde{\rho}\, (o)/(c)}_e  =& 2.61 \mu  \sqrt{\frac{2 r-1}{r^5}} \Omega  \cot \theta \,, \notag \\
B^{\tilde{z}\, (o)/(c)}_e  =& 1.74 \mu  \sqrt{\frac{2 r-1}{r^5}} \Omega  \left(\cot ^2\theta -\frac{1}{2}\right) \,, 
\end{align}
while the expression for $B^{\tilde{\phi}}_e$ in the cylindrical co-ordinates is still given by Eq.~\eqref{eq:Bphi}. 

%\begin{figure}[tb] 
%  \centering
%\begin{overpic}[width=0.8\columnwidth]{TopViewCS.pdf}
%\end{overpic}
%  \caption{The direction of the LOFLs as viewed from above (the magnetic axis points out of the paper). The red circle marks the location of the light cylinder.}
%       \label{fig:TopView}
%\end{figure}

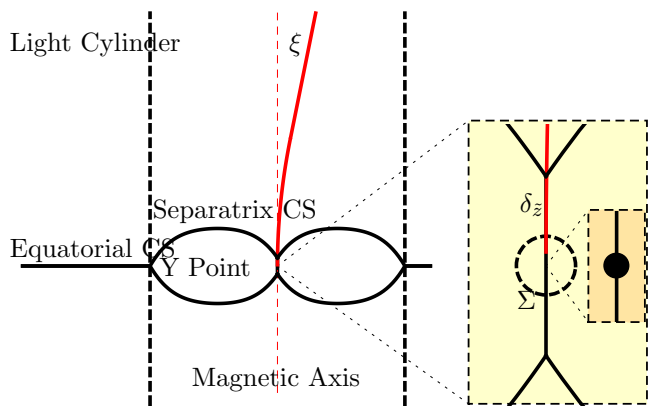
\begin{figure}[tb] 
  \centering
\begin{overpic}[width=0.95\columnwidth]{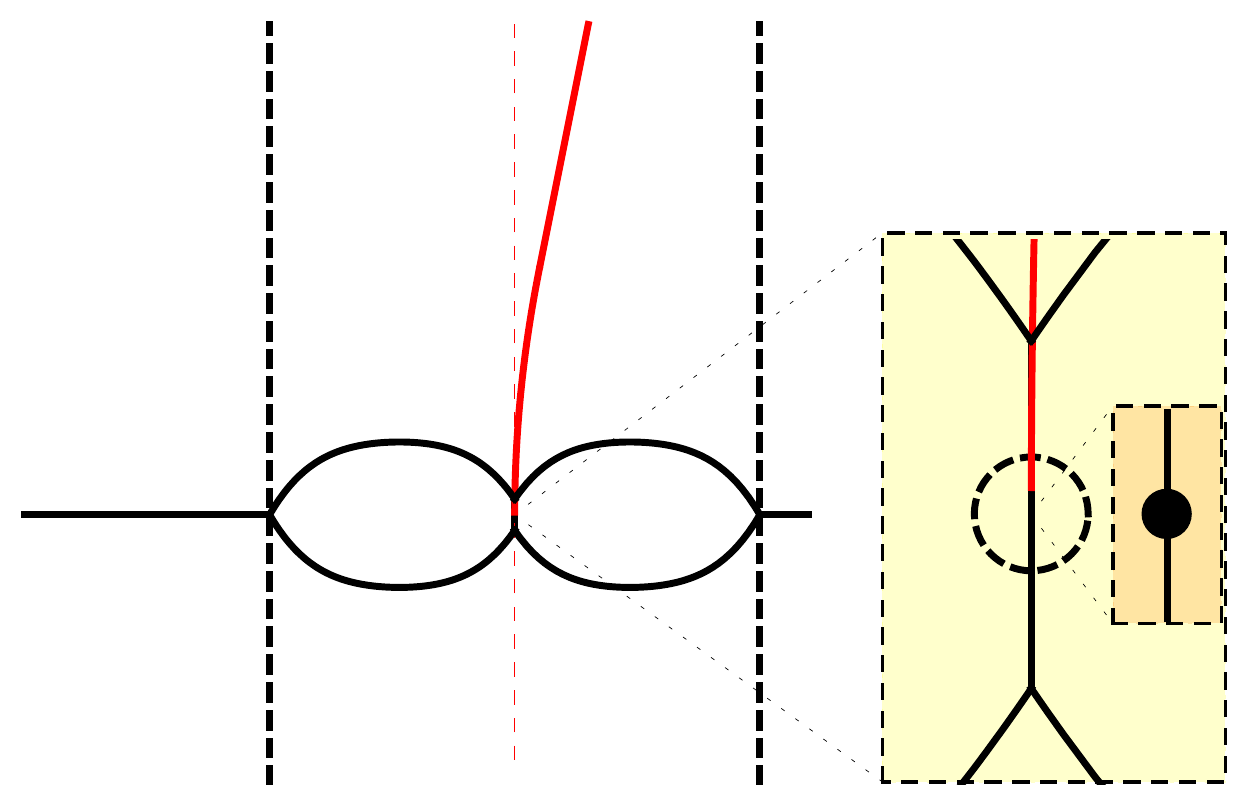}
\put(43,55){$\xi$}
\put(0,23){Equatorial CS}
\put(23,20){Y Point}
\put(22,29){Separatrix CS}
\put(28,3){Magnetic Axis}
\put(0,55){Light Cylinder}
\put(78.5,30){$\delta_{\tilde{z}}$}
\put(78,15){$\Sigma$}
\end{overpic}
  \caption{Poloidal (constant $\tilde{\phi}$) slice of Fig.~\ref{fig:CSPic} to show the relative scales of the various quantities. The CSs are shown as thick solid black lines, the light cylinder as two dashed vertical lines, and a radio signal path is shown as a red solid curve. 
The two layers of insets zoom in on the central parts with increasing levels of magnification. We note the vertically rising CS segments in the insets are artefacts of the polynomial fit \eqref{eq:CSProfile} being too rigid to account for the rapid rise in altitude of the CSs there, the real CSs are more smooth. The dashed circle $\Sigma$ signifies the upper altitudinal limit by which the transition from Alfv\'en to isotropic waves must be accomplished, so the light path does not need to track the LOFLs/CS outside of it. The solid black disk represents the neutron star. }
        \label{fig:Sketch2}
\end{figure}

Since in the inner core regions of interest to us (still much larger than the stellar volume, see Fig.~\ref{fig:Sketch2} for scales), the magnetic field lines are almost entirely poloidal, there would be hardly any Alf\'en waves emitted into toroidal directions. Instead, they travel outwards away from the neutron star (within the separatrix CSs, the current flows inwards and the negatively charged electrons stream outwards, as do the waves they excite) in directions initially (at low altitudes) closely hugging the magnetic axis (we denote the angle between the propagation directions and this axis by $\xi$). One can glean a sense of the closeness (the smallness of $\xi$) by noting that the size of the neutron star is minuscule as compared to the scales in the magnetosphere (see Fig.~\ref{fig:Sketch2}), so the CSs can, on intermediate scales to leading order approximation (see footnote \ref{ft:Dipole}), be seen as following a subset of field lines of a point-like magnetic dipole, which must strike the infinitesimal source point along the magnetic axis (corresponding to $\xi =0$). 
The small size of the opening angle in fact results in the fitted poloidal profile \eqref{eq:CSProfile} of the CSs having an offset $\delta_{\tilde{z}} \approx 4000$km at $\tilde{\rho}/R_{\rm LC} \sim 1/R_{\rm LC}$  (stellar surface is at $\tilde{\rho}^2+\tilde{z}^2 \approx 1$), which is of course due to our polynomial fitting not being flexible enough to properly track the almost vertically rising $\tilde{z}$ at very small $\tilde{\rho}$, but nevertheless provides a convenient approximate delimiter for when the CSs begin to noticeably broaden. In comparison, the transition from Alfv\'en into isotropic waves needs to be accomplished at altitudes below $1000$km, in order to be consistent with the lack of Landau absorption signatures in the observed radio signal spectra \citep{1986ApJ...302..120A}. Since the offset $\delta_{\tilde{z}}$ is larger than this threshold, the natal Alfv\'en waves will not follow the CSs all the way out to their more opened-up regions. There will however be refraction effects that slightly increase $\xi$, given that $\eta^{(2)}$ drops as the waves rise in the magnetosphere. A more quantitative assessment of $\xi$ impinge on the availability of precise knowledge of the plasma density and magnetic field distributions within the magnetosphere, but the general picture (Fig.~\ref{fig:Sketch2}) appears to be in agreement with the observed radio signals from pulsars being quite narrow in temporal duration (e.g. as compared to high frequency emissions). For example, the first pulsar discovered, PSR B1919+21, has a period of $1.34$s but a pulse width of only around $0.04$s, translating into a beam opening angle about one tenth of the pulsar inclination angle.

\section{Comparison with observations} \label{sec:Obs}
Because linearly polarized waves propagating along a hollow cone matches the basic geometric or phenomenological model for radio emission inferred from observations \citep{1970Natur.225..612K,2015SSRv..191..207B}, the CSCR mechanism is consequently able to account for the most rudimentary features of pulsar radio signals, such as double peaks when the telescope makes central passages through the beam (curve A in Fig.~\ref{fig:Distributivity} below) and single peaks for grazing passages (curve B), together with an S-shaped swing in the polarization position angle \citep{1969ApL.....3..225R}. 
On the other hand, drivers behind the more nuanced details are more difficult to ascertain, because there seems to always be exceptions to any rule. Nevertheless, it is worthwhile taking a closer look at those features of some generality. 

\subsection{Frequency space profiles} \label{sec:Freq}
\begin{figure}[tb] 
  \centering
\begin{overpic}[width=0.95\columnwidth]{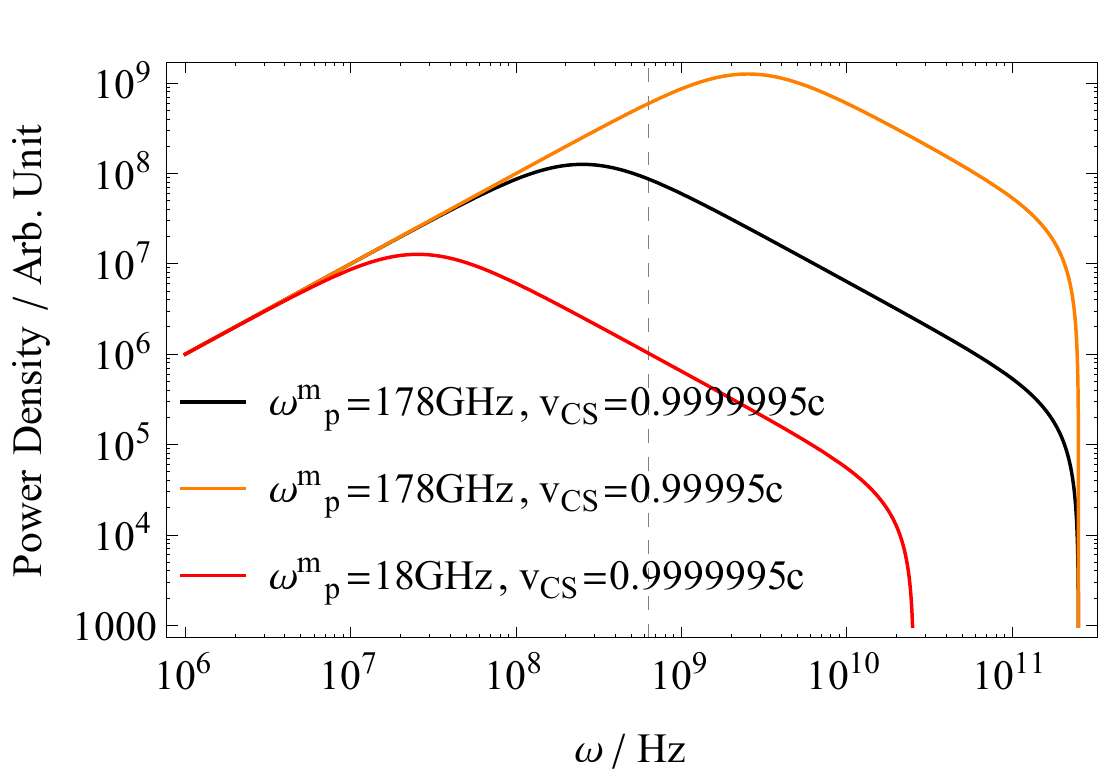}
\end{overpic}
  \caption{Power density spectrum at emission site as given by Eq.~\eqref{eq:Knee}, for a variety of magnetospheric parameters. The black curve corresponds to our fiducial parameter values in Table~\ref{tb:Params}. The effect of the cutoff (the precipitous drop on the far right) and the existence of a break frequency (the peak near $2\pi \times 100$MHz, marked with a vertical dashed line) are clearly visible.}
        \label{fig:SourcePower}
\end{figure}

{The more salient features that can help distinguish between different emission mechanisms are often found within the frequency space profiles of the observed radio signals, that is, in the flux density spectra. We show here that the CSCR model can produce the correct spectral shape, including the low-frequency turnover in particular}. We begin by 
denoting with $\omega_p^m$ the $\omega_p$ value at emission site, then jointly solving Eqs.~\eqref{eq:Direction} and \eqref{eq:Propagation}, we obtain
\bea\label{eq:ThetaSol}
\theta = \arctan\left[\sqrt{1-\frac{v_{\rm CS}^2}{c^2}}\sqrt{2 (\omega^m_p)^2/\omega^2-1}\right]\,,
\eea
which admits real solutions only when $\omega<\omega^m\equiv \sqrt{2}\omega^m_p$ (see also Fig.~\ref{fig:Eta2Sqr})\footnote{We also note that there is no explicit lower limit on $v_{\rm CS}$ in this expression because $\eta^{(2)}$ can diverge at certain angles (see Fig.~\ref{fig:zeta} (a)) making Eq.~\eqref{eq:Direction} easier to satisfy. In reality though, additional complications such as collisions will regularize the divergence and place a constraint on $v_{\rm CS}$ as well.}. The fiducial parameters in Table~\ref{tb:Params} predict the cutoff frequency at $\omega^m/(2\pi) \sim 40$GHz (the observational frequencies are not normally quoted as angular frequencies, thus the $2\pi$ factor), while the actual highest radio frequency at which an observation has been achieved lands at a very similar $87$GHz \citep{1997A&A...322L..17M}. 
Below the cutoff, Eq.~\eqref{eq:ThetaSol} allows us to evaluate the detailed \u{C}erenkov radiation power density spectrum as given by the Frank-Tamm formula (see e.g. \citet{1955BJAP....6..227J}). Substituting in Eqs.~\eqref{eq:Direction} and \eqref{eq:ThetaSol}, we have that the power spectrum at the emission site is proportional to 
\begin{align} \label{eq:Knee}
\left(1-\frac{c^2}{v^2_{\rm CS} \eta^2} \right)\omega =&
\left(1-\cos^2\theta\right)\omega 
\notag \\
=& \omega -\frac{c^2 \omega^3 }{2 c^2 (\omega^m_p)^2+v_{\rm CS}^2 \left(\omega^2-2 (\omega^m_p)^2\right)}\,.
\end{align}
We have plotted expression \eqref{eq:Knee} in Fig.~\ref{fig:SourcePower}, which exhibits a turn-around ($\theta$ rises rapidly towards $\pi/2$ there, as Fig.~\ref{fig:Direction} also shows) centred on a break frequency at 
\bea
\omega_{b} = \sqrt{2}\omega_p^m \sqrt{\frac{c^2}{v_{\rm CS}^2}-1}\,,
\eea
where the $\omega$ and $\omega_p$ contributions to the denominator in Eq.~\eqref{eq:Knee} become comparable. This location of the turn-around can be adjusted by tuning either $\omega_p$ or $v_{\rm CS}$, and falls on $\omega_b/(2\pi)\approx 40$MHz for our fiducial parameters in Table~\ref{tb:Params}. 

\begin{figure}[tb] 
  \centering
\begin{overpic}[width=0.95\columnwidth]{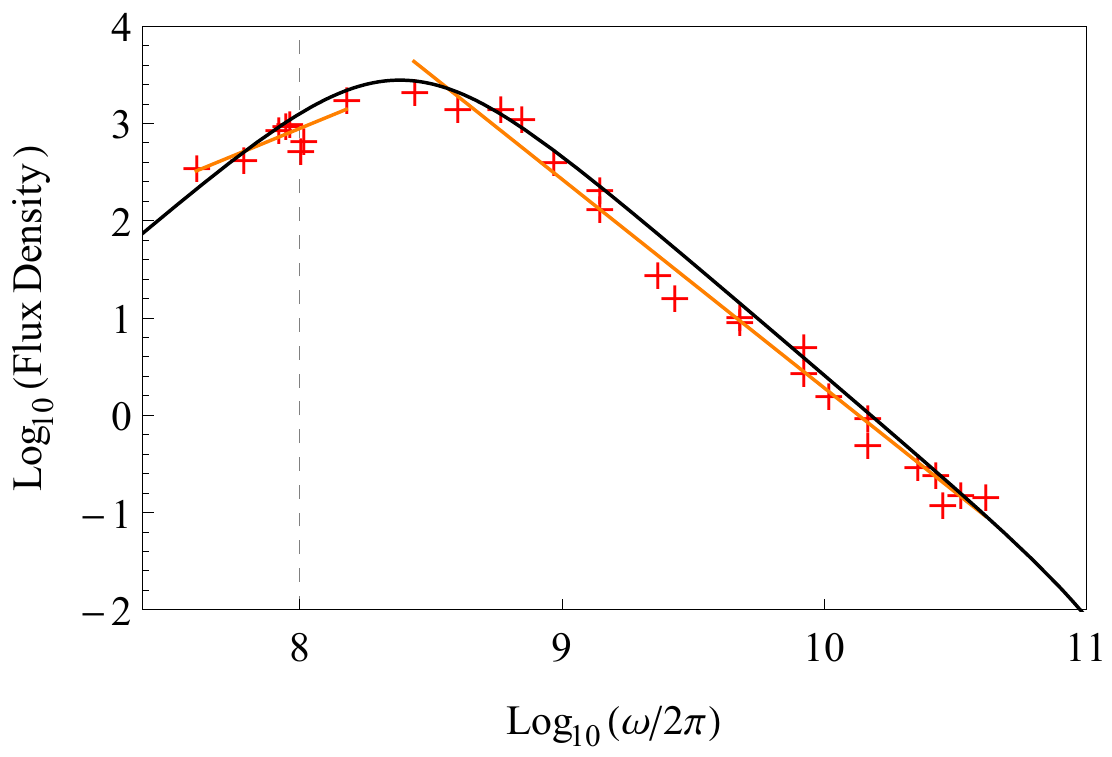}
\end{overpic}
  \caption{Example flux density spectrum for PSR B0329+54 that exhibits a low-frequency turnover. The data are taken from \citet{2012hpa..book.....L} and plotted as crosses. The straight lines are the best linear fits for segments of the spectra, with slopes of $+1.1$ and $-2.1$. The black curve is a fit using Eq.~\eqref{eq:Knee} with $v_{\rm CS}=0.9999995c$ and $\omega^m_p=10^{12}$Hz. This is not the only parameter set to achieve a fit, offsetting adjustments in $v_{\rm CS}$ and $\omega_p$ can be applied while still maintaining the fit.}
        \label{fig:Spectrum}
\end{figure}

The breaking of the curves in Fig.~\ref{fig:SourcePower} into two regimes visually resembles the so-called low frequency turnover often seen at around $100$MHz \citep{1996ASPC..105..271M}, and we can in fact achieve rather good quantitative fits for for example, PSR B0329+54 (used as an example in the standard pulsar textbook \citet{2012hpa..book.....L}), by simply starting from the fiducial parameters in Table~\ref{tab:params} and increase $\omega_p$ by a factor of six (or alternatively keeping $\omega_p$ the same but change $v_{\rm CS}$ to $0.999985$c). The `fitting' result is shown in Fig.~\ref{fig:Spectrum}, and we note that although $\omega_b$ is adjustable (we still would like to stay close to the fiducial parameters, so the range for even this quantity is constrained), the slopes of the two regimes separated by it are essentially fixed (as demonstrated by Fig.~\ref{fig:SourcePower}; thus the quotation mark surrounding the word fitting), yet the CSCR predictions for them turn out to match observations remarkably well. 

Looking through larger sample sizes, the spectra for different pulsars exhibit a rather wide variety. Although power laws are quite generically applicable, the power index $\alpha$ for the segment above $\omega_b$ can range from as high as $\alpha\sim 0$ to as low as $\alpha\sim -4$, averaging though at $-1.8\pm 0.2$ \citep{2000A&AS..147..195M}, which is similar to the value for PSR B0329+54. A possible explanation for this spread arises from the observation that higher frequency waves tend to correspond to smaller beam opening angles (narrower pulse width) \citep{2016JPlPh..82b6302M,2012hpa..book.....L,2014ApJS..215...11C}, causing a variation in the flux distributivity across the beam amongst different frequencies. Depending on our line-of-sight, we may graze either increasingly peripheral (if the passage through the beam, cf. Fig.~\ref{fig:Distributivity}, is on the outer rim of the hollow cone) or more central regions (if inner rim) of the frequency-specific constituents of the radio beam as we move up in $\omega$, observing suppressed or enhanced (or fortuitously little impacted when in-between, as is perhaps the case with PSR B0329+54) intensity at higher frequencies as a result (this is an additional geometric effect that needs to be multiplied onto the power spectrum at source given by Eq.~\eqref{eq:Knee}). This frequency dependence of the opening angles can be understood in the CSCR context by examining Fig.~\ref{fig:Eta2Sqr}, which shows that the transition from the LOFL-following Alfv\'en waves to freely propagating isotropic waves occur at lower $\omega_p$ (higher altitude, so the opening angle would have broadened more during the LOFL-tracking phase prior to the transmutation) for smaller $\theta$s, which correspond to higher frequencies. 

{Finally, we note that although general relativistic computations are invoked to compute the magnetic field strength in previous sections, it is really an overkill (inherited from the foundational reference that our computations are based on). For the emission region of the order of $1000$km above the star surface (the Schwarzschild radius of the star is around $4$km in comparison), the gravitational redshift alters the frequency of the emitted radio signals by around $0.2\%$, and is negligible at the accuracy with which we can measure the spectra. }

\subsection{Temporal profiles} \label{sec:TempProf}
{We turn next to the temporal profiles of the observed radio signals, which often consist of sub-pulses (that are themselves further decomposable into micro-pulses) that drift across the mean pulse window. These features are more reflective of the geometry of the emission region (the distribution of the emitting sources), and are as such not uniquely associated with \u{C}erenkov radiation. Nevertheless, the temporal profiles are quite rich in structure, which incidentally agree with certain properties that one would expect of the CSs. Therefore, they still possess some distinguishing power since the emitting sources being the CSs is a central and rather unique (among coherent radio emission models) feature of the CSCR model.} 

Specifically, thin CSs are prone to developing plasmoids in the $\tilde{\phi}$ direction transverse to current flow (parallel currents tend to repel each other) \citep{2015bps..book.....C}. This transverse instability (in contrast, the dynamics along the current flow direction is stable, thanks to the dissipation through \u{C}erenkov radiation reaction, see Sect.~\ref{sec:CS}) produces islands of magnetic field lines more tightly bundled together, but will not choke off the current flow via the mirror effect, since the synchrotron loss time is less than $3\times 10^{-14}(\bar{\gamma}_{\rm CS}/100)(10^8T/B)^2$s \citep{1986ApJ...302..120A}, which is far smaller than any other dynamical timescales near the emission sites, so the electrons `instantly' adapt their pitch angles to go through the bottlenecks. The bundling will however concentrate the current, and thus the islands correspond to emission sites that produce more intense \u{C}erenkov radiation. Furthermore, because the Amp\`ere's forces and the magnetic pressure have components along $\tilde{\phi}$ (squeezing on a flux tube will also upset the force balance for the neighbouring field lines), the magnetic field line displacement should propagate as a longitudinal wave along this direction. Because there exists influences such as the Coriolis forces (acting on the fast streaming electrons in the CSs) that break the symmetry between clockwise and anticlockwise propagations, we expect their associated wave speeds to be slightly different. On the other hand, because these current and magnetic flux density waves moving in either direction need to satisfy the same cyclic boundary conditions, their wavelengths must be equal, so there must be small differences between their angular velocities $\omega_{c}$ (subscript $c$ for clockwise) and $\omega_{a}$ ($a$ for anticlockwise).  

\begin{figure}[tb] 
  \centering
\begin{overpic}[width=0.8\columnwidth]{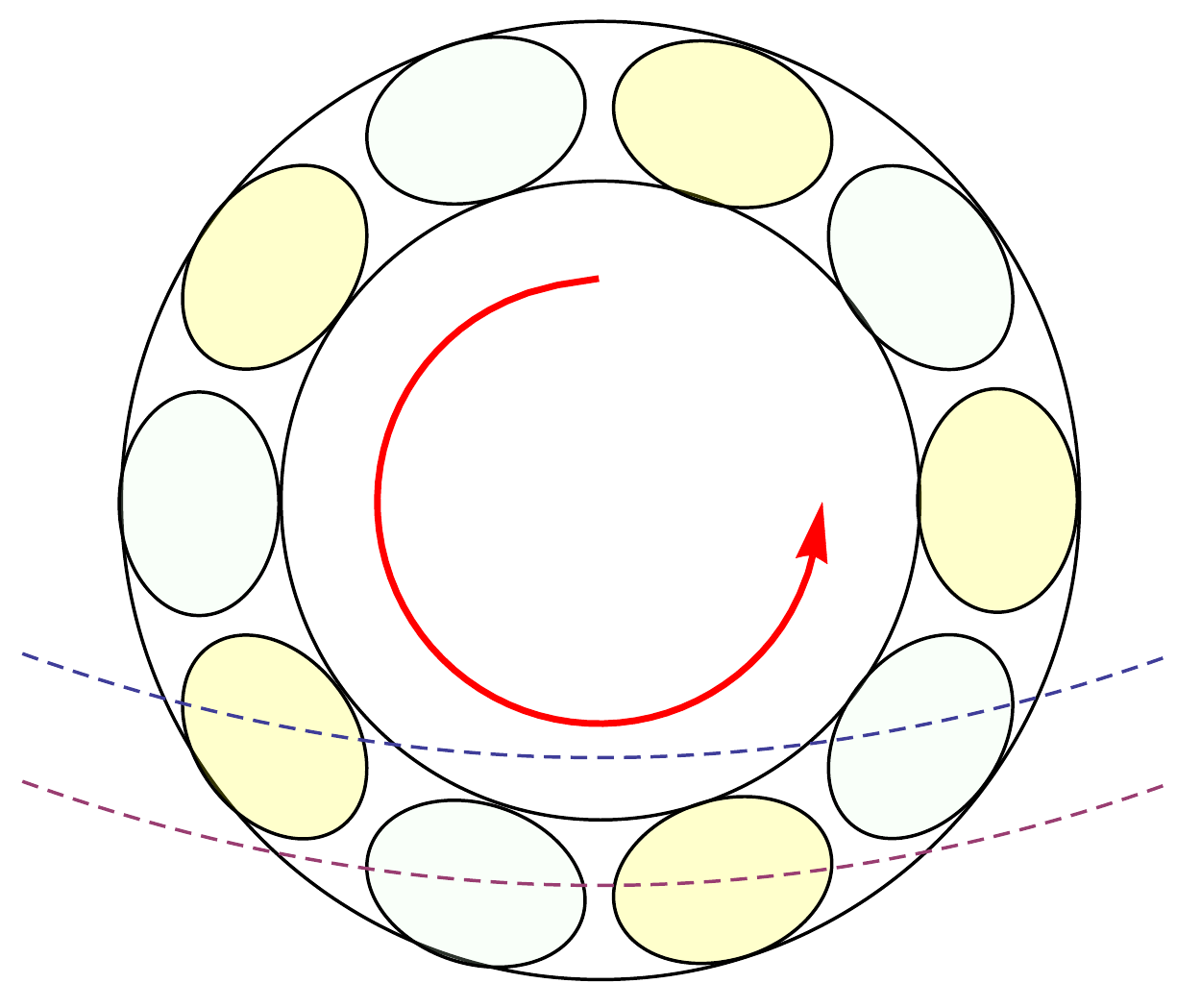}
\put(2,24){A}
\put(2,13){B}
\end{overpic}
\begin{overpic}[width=0.99\columnwidth]{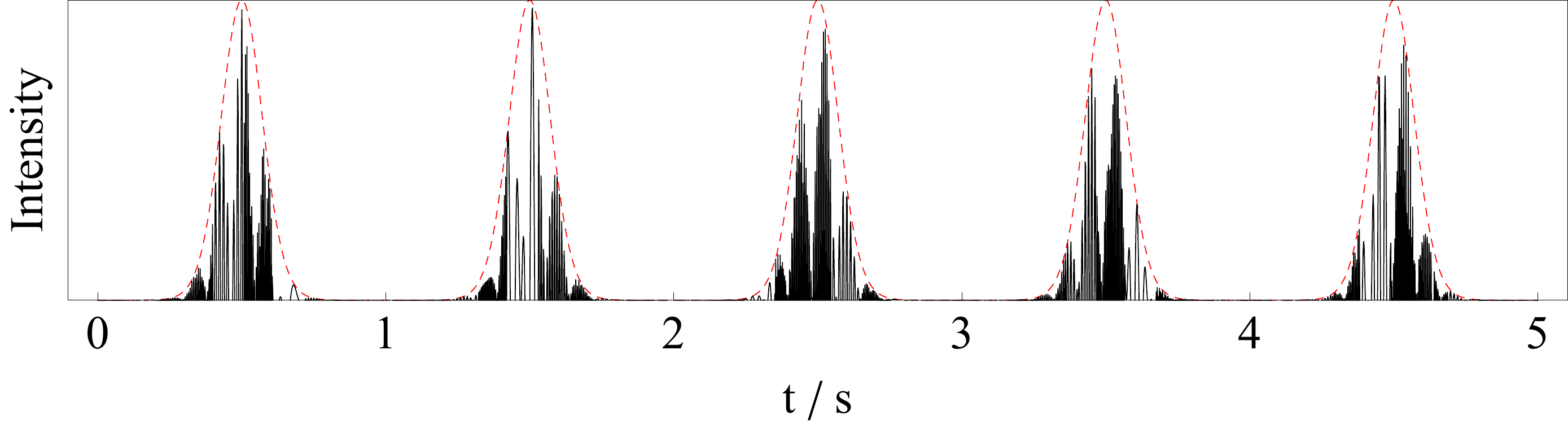}
\end{overpic}
  \caption{Top: Schematic drawing of the distributivity pattern across a radio beam (looking down the magnetic axis). The annulus represents a cross-section of the emission beam, while the two dashed curves represent the telescope's paths traversing this cross-section. The yellow and light-green shaded ellipses represent the peaks and troughs of the beating pattern formed from two counter-circulating current density waves, and physically correspond to the sub-pulses (note they are indeed observed to be almost evenly spaced, see e.g. \citet{1981IAUS...95..199B}), with yellow corresponding to $\delta\Phi>0$. They rotate slowly around the magnetic axis as indicated by the red arrow, while at the same time rapidly switch colour. Bottom: Simulated signal using Eq.~\eqref{eq:Beating} overladen with an envelop (red dashed line) representing the effect of making a B passage (see top panel) periodically. The parameters used are slightly adjusted as compared to the values given in the main text to make the signals wider and their features more visible in the figure. The jaggedness is due to the limited number of sampling points used by the plotting software, and conveniently simulate the limited sampling rate of a telescope. The drifting of the sub-pulses is visible as the tip of the red envelop starting off coincident with a sub-pulse, but shifting to between two sub-pulses at the end of the sequence. Only $\delta\Phi>0$ segments are shown, as in real signals the dim $\delta\Phi<0$ segments are usually overwhelmed by noise.}
        \label{fig:Distributivity}
\end{figure}

Assuming first that one particular angular wave number $\hat{m}\in \mathbb{Z}$ dominates, then a quick illustrative calculation for the magnetic flux density perturbation
\begin{align} \label{eq:Beating}
\delta\Phi(t,\tilde{\phi})=&\cos\left(\omega_{c} t + \hat{m}\tilde{\phi}\right)+\cos\left(\omega_{a} t-\hat{m}\tilde{\phi}\right)\notag \\
=&2\cos\left[\left(\frac{\omega_{c}+\omega_{a}}{2}\right)t\right]\cos\left[\left(\frac{\omega_{c}-\omega_{a}}{2}\right) t + \hat{m}\tilde{\phi}\right]
\end{align}
shows that we would have an imperfect standing wave with a beating pattern (the second multiplicative factor in Eq.~\ref{eq:Beating}) slowly drifting with an effective period $4\pi/(\hat{m}(\omega_c-\omega_a))$. 
We have thus, via a very different mechanism, reproduced the phenomenological model of a carousal of sub-beams within the hollow emission cone (see Fig.~\ref{fig:Distributivity}), as proposed by \citet{Ruderman:1975ju}, which has been very successful at explaining the sub-pulse drifts \citep{1999ApJ...524.1008D,2002A&A...393..733E}. Our present proposal differs from the original in finer phenomenological details however, particularly in that as the first multiplicative factor in Eq.~\eqref{eq:Beating} rapidly oscillates between being positive and negative, the peaks and troughs of the second factor alternate in lighting up in quick succession ($\delta\Phi>0$ regions correspond to higher current density and are thus brighter than average). Assume that the clockwise and anticlockwise current density wave speeds are both close to the speed of light (they need only differ of the order of $0.0004\%$ to produce the sub-pulse drifting period that's typically ten times the fundamental pulsar period), that a $\tilde{\phi}$ cycle covers of the order of {$600$km (circumference of a circle of radius $100$km)} at the emission altitude of the order of $1000$km (corresponding to a beam opening angle of the order of $1/10$, see Fig.~\ref{fig:Sketch2}), and $\hat{m}$ is of the order of ten (so we catch a couple of sub-pulses with each crossing of the beam, see Fig.~\ref{fig:Distributivity}), then $\omega_{a/c} \sim 3\times 10^4\text{s}^{-1}$, corresponding to a period of $200\mu$s. This is smaller than the duration of a sub-pulse (typically of the order of $10$ms), so as we make our way through any shaded ellipsis in Fig.~\ref{fig:Distributivity}, it would flash between bright and dim states (i.e. switch colour) many times. In other words, each sub-pulse is in fact further decomposed into constituent peaks that are hundreds of microseconds wide. This provides an explanation for the quasi-periodic micro-pulses seen sitting on top of the sub-pulses \citep{1968Natur.218.1122C,1971ApJ...169..487H} (see the bottom panel of Fig.~\ref{fig:Distributivity} for the hierarchy of predicted pulses, sub-pulses and micro-pulses), which indeed have the predicted width (see e.g. Table~2 in \citet{2002MNRAS.334..523K}). 

As an interesting aside, we also note that the drifting pauses during pulsar nulling periods and simply resumes from where it left off afterwards \citep{1973MNRAS.163...29P}. This could be understood as plasma depletion attenuating the number of electrons flowing away from the star along CSs, leaving magnetic pressure dominating the restoring forces for the flux density wave, thus we have $\omega_{c}=\omega_{a}$ and the beating pattern freezes into a true standing wave, to be reanimated when the current flow returns to normal levels. Plasma replenishment is however a more gradual process, thus we observe the characteristic transient dip in the sub-pulse drift rate immediately after each nulling (see Fig.~4 (b) in \citet{1973MNRAS.163...29P}).

As more $\hat{m}$ ingredients (all must be integers) are added for some pulsars, the sub-pulses become increasingly complicated in their distribution pattern across the mean pulse profile, as well as in their apparent drifting behaviour. Eventually, clear trends become difficult to discern, and the pulse structure evolution appears erratic. In this limit, we essentially recover a hollow cone version of the patchy core phenomenological model of \citet{1988MNRAS.234..477L}.

\section{Conclusion} \label{sec:Conc}
The pulsar magnetosphere hosts many processes operating on varying time scales, usually in a hierarchy different from what we encounter in terrestrial laboratories. Such oddities allow for new possibilities, and our admitted crude estimates show that the pulsar environment may be conducive to \u{C}erenkov radiation by fast streaming electrons flowing along the separatrix CSs, and that features relating either to the source particle motion or the emission process itself could yield semi-quantitative agreements to the most salient features of the pulsar radio signals, in both temporal and frequency domains. 

{When compared with alternative models already existing in literature, the CSCR model enjoys the benefit of the source particles engaging in an orderly motion in the poloidal direction, yet simultaneously the CSs can develop fine structures in the toroidal direction (see Sect.~\ref{sec:TempProf}). The  collective poloidal motion quite naturally takes care of the extremely high brightness temperature of $10^{25}$K to $10^{30}$K \citep{1975ARA&A..13..511G} or in excess of $10^{40}$K for the Crab pulsar \citep{2007ApJ...670..693H} (it is not practical to actually achieve such temperatures, so disorderly thermal emission is generally not considered viable, and coherent emission mechanisms such as the one investigated here had been the focus of model-building exercises). On the other hand, the toroidal fine structures can account for the nuanced features in the temporal signal profiles, translating into length scales as small as several metres. }

% collective motion of CS electrons means they are more like particles with large q, so Frank-Tamm being propto q^2 means N particles will radiate N^2 times the power, just like the bunching.

{The existence of such fine structures have perhaps helped motivate modelling efforts that invoke instabilities (a commonality among antenna, plasma emission and maser mechanisms), in the form of extremely bright coherent emissions from small regions hosting instabilities that then organize themselves into more orderly larger scale structures (we note however that small length scales are really only required for the toroidal or lateral direction of the emitting structure, as this is the direction that we sample when the radio beam sweeps across Earth). For example, self-bunching beam instabilities are proposed in which bunches of $N$ particles (moving in unison, behaving like one particle with $N$ times the charge) form in small volumes which then radiate at $N^2$ (power is proportional to charge squared) rather than $N$ times (as in the incoherent case) the individual particle power, during, for example, curvature emission (see e.g. \citet{Ruderman:1975ju}, and also \citet{1998A&A...332L..21L} for the inconsistency of this model with observed luminosities, as well as \citet{1992msem.coll..306M} for a discussion on the coherence ceasing to be effective when the bunches are of finite sizes). A common difficulty encountered by instability-based models is that the extremely high brightness temperatures require very rapid growth rates of the instabilities (maintaining coherence within possibly turbulent environments engendered by such violent instabilities could pose further challenges to all instability-based models), which are not easily achievable (e.g. this has been a problem for some maser models \citep{2016JPlPh..82b6302M}) and often demand resonances (with instabilities, resonances generically exist even if they are not absolute prerequisites for radio emission, see e.g. \citet{2002MNRAS.337..422G} in the context of plasma emissions). However, there are no obvious signs of narrow resonance peaks in the observed flux density spectra (even if the peaks are to be broadened by variations in the magnetospheric conditions, the spectra still appear arguably too smooth). The resonance conditions may also depend sensitively on the magnetic field strength (see e.g. \citet{1991AuJPh..44..573K}), excluding such models from simultaneously explaining radio emission from different classes of pulsars. Moreover, the back-reaction of the emission tends to suppress instabilities \citep{2016JPlPh..82b6302M}.}

{In contrast, the CSCR model is not dependent upon any in situ growing instability, that needs to be sustained for sufficient durations to achieve the necessary brightness (the mild plasmoids in the toroidal direction are an incidental feature that is not a required ingredient of the radiation mechanism; its absence would not shut down the \u{C}erenkov radiation). The energy budget for the sourcing electrons is instead acquired elsewhere upstream via steadily ongoing processes, but only released when the conditions are suitable for \u{C}erenkov radiation, which incidentally is broad-band, exhibiting no sharp resonance peaks (the Frank-Tamm formula is a simple power law for a non-dispersive medium, and is only a little more complicated in pulsar magnetospheres, please see Sect.~\ref{sec:Freq}). The flip side of this `non-locality' is that the CSCR model is highly dependent on the global structure of the magnetospheres, which might be subject to revision in the future. 
Another drawback of the CSCR model is that there are no similar processes from for example, solar physics, that we have become familiar with, and it is difficult to reproduce the extreme conditions on Earth, so we have to rely on theoretical (including numerical in particular) arguments to further ascertain its viability, or lack of. Yet the treatment of resistivity in the bulk of pulsar magnetospheres and the handling of the CSs themselves remain rather rudimentary at the moment (e.g. many force-free simulations simply truncate electric fields when they exceed the magnetic field in amplitude \citep{2015PhRvD..92b4049Z}).}

In addition to these difficulties, to keep the scope of the paper contained and the discussion focused, we have deliberately avoided straying too far from the fiducial parameters relevant for a normal pulsar. It is nevertheless interesting to note that while in most cases the higher frequency radiations do not share the profile and phase of radio signals \citep{2005tsra.conf..149H,2013ApJS..208...17A,2016ApJ...833...47H}, strong similarities do occur for young pulsars such as the Crab \citep{1996ApJ...468..779M,2016ApJ...833...47H}, and correlation is also seen in millisecond pulsars \citep{2003A&A...410L...9C,2003ApJ...597.1049K}. Assuming the high frequency emissions to indeed originate from magnetic reconnections, then having radio signals to also be sourced by the CSs simplifies efforts to explain such phenomena. Namely, with either the very young or the recycled millisecond pulsars, the light cylinder and the \verb!Y! point (see Fig.~\ref{fig:Sketch2}, this is an active spot for reconnections \citep{2015MNRAS.448..606C}) shift very close to the star ($\sim 1500$km for a $30$ms period, and scales linearly with the period), so the reconnection and \u{C}erenkov radiation regions, both confined to the shrunken CSs, may simply overlap. 

With the new and more sensitive telescopes such as the Five-hundred-metre Aperture Spherical radio Telescope (FAST) \citep{2006ChJAS...6b.304N,2009A&A...505..919S} and the Square Kilometer Array (SKA) \citep{2015aska.confE..36K}, we expect more exquisite structural details of the radio signals to be unveiled for more pulsars, allowing to study their variation against magnetospheric parameters, thereby helping to either verify or falsify the CSCR model. If turns out to be on the right track, we expect this model to be useful in understanding the jitter noise that limits the precision of pulsar timing arrays and thus their sensitivity to gravitational waves. 

\begin{acknowledgements}
The author thanks Di Li and George Hobbs for helpful discussions on jitter noise, which motivated the present study. {We also thank an anonymous referee for insightful comments pointing out many subtleties, as well as their many helpful suggestions, which helped to substantially improve the clarity and completeness of the paper.}
This work is supported by the Strategic Priority Research Program of the Chinese Academy of Sciences Grant No. XDB23000000, the
National Natural Science Foundation of China grants 11503003 and 11633001, the Fundamental Research Funds for the Central Universities grant 2015KJJCB06, and a Returned Overseas Chinese Scholars Foundation grant. 
\end{acknowledgements}

\bibliographystyle{aa} % style aa.bst
%\bibliography{../../../../apj-jour,../../../../references}
\bibliography{paper.bbl}

\end{document}